\documentstyle[epsf]{mn}
\begin{document}

\title[Period-magnitude relations for M giants]{Period-magnitude
relations for M giants in Baade's Window NGC\,6522}

\author[I.S. Glass, M. Schultheis]{I.S. Glass$^1$ and M. Schultheis$^{2,3}$ \\
$^1$South African Astronomical Observatory, PO Box 9, Observatory 7935, South
Africa\\$^2$Institut d'Astrophysique, 98 bis Blvd Arago, Paris,
France\\$^3$Observatoire de Paris-Meudon, GEPI, Place Jules Janssen
2, F-92195 Meudon Cedex, France }

\date{Submitted 2003;
accepted}

\maketitle

\begin{abstract}

A large and complete sample of stars with $K$ $<$ 9.75 in the NGC\,6522
Baade's Window towards the galactic bulge is examined using light curves
extracted from MACHO and $IJK$ photometry from DENIS.

The improved statistics resulting from a sample of over 1000 variables allow
all four of the sequences A, B, C and D in the $K_S$, log$P$ diagram of the
Large Magellanic Cloud to be seen in the Bulge. The Bulge sequences,
however, show some differences from those in the Magellanic Clouds, possibly
due to the effects of higher metallicity. These sequences may have value as
distance indicators.

A new diagram of the frequency of late-type variables at a given amplitude
is derived and compared to the old one due to Payne-Gaposchkin (1951). The
catalogued semi-regular variables of the solar neighbourhood are found to
be a subset of the total of semi-regulars, biased towards large amplitude.
 
\end{abstract}


\section{Introduction}

The existence of period-luminosity relations is of great importance to
astronomy. Each class of variable that shows such a relation adds to the
precision with which distance comparisons can be made between nearby
galaxies and ultimately assists in the determination of the distance scale
of the universe. Clearly, the limits of precision and the dependence of a
relation on factors other than distance must be understood if it is to be
truly useful. Among the late-type variables it has been known for over twenty
years that the Miras obey a good $K$, log $P$ law in the infrared, where they
emit most of their energy (e.g., Glass \& Lloyd Evans, 1981, 2003). More
recently, large-scale variability surveys such as MACHO (Alcock et al,
1999) and the near infrared sky surveys DENIS (Epchtein 1998) and 2MASS
(Skrutskie 1998) are leading to new insights into the behaviour of the
semi-regular variables (SRVs), which require long runs of data on many
objects if their systematic and general properties are to be elucidated.
Within the Milky Way galaxy, the Baade's Windows towards the inner Bulge are
particularly suitable for studies of this kind because most of the objects
they contain are at an approximately uniform distance that is known with
moderate precision.

Alard et al (2001; see also Glass \& Alves, 2000) found that nearly all the
stars detected during the infrared ISOGAL survey (Omont et al, 2003) at 7
and 15 $\mu$m of the Baade's Windows NGC\,6522 and SgrI are, in fact,
late-type variables, and that many of those with periods in excess of 80
days are losing mass. Such stars contribute significant amounts of re-cycled
material to the interstellar medium. By identifying the Alard et al stars in
the DENIS $IJK_S$ data, Schultheis \& Glass (2001) have been able to
investigate their general properties in the near-infrared colour-magnitude,
colour-colour and colour-period diagrams. They have also been
cross-identified with a complete sample of M giants with spectral
classification by Blanco (Glass \& Schultheis, 2002; GS) and the dependence
of their variability properties on spectral type and $K$ magnitude has been
determined. In general, it was found that significant variability occurs
only amongst Blanco subtypes of M5 or later (corresponding to M-K subtypes
M4 or later).

Wood and the MACHO team (1999) and Wood (2000) showed that the AGB stars in
the Large Magellanic Cloud (LMC) fall into four approximately parallel
sequences (one of which corresponds to the Miras) in the $I_W$ vs log $P$
and $K$ vs log $P$ diagrams. Although our Baade's Window work suggested that
there was some evidence that the Bulge giants might behave similarly, the
numbers of stars were too few to demonstrate this at all convincingly and it
was also suspected that the depth of the Bulge along the line of sight was
smearing out any systematic trend. Similar results are emerging from studies
of the Small Magellanic Cloud (SMC, see Cioni et al and Ita et al, in
preparation)

In addition, recent evidence from the LMC (Cioni et al, 2001) suggests that
there is more continuity in the amplitude distribution with period of the
late-type variables than has hitherto been believed. The canonical view of
this distribution rests on the work of Payne-Gaposchkin (1951), based on
solar neighbourhood studies made with visual and photographic techniques and
with no certainty as to the uniformity of the sample in space. From
Payne-Gaposchkin's work it appeared that a gap existed in the amplitude
distribution between the Miras and the irregular variables. This could have
been an artefact arising from the ease of detecting Miras in spite of their
rarity and the probably similar rarity of stars with intermediate
amplitudes, which are much harder to find. In addition, small-amplitude
variables will have been under-represented in her data because of the
limitations of photographic photometry.

These two considerations suggested that a large-scale investigation of the
AGB variables in the NGC\,6522 Baade's Window as they appear in DENIS and
MACHO would be worthwhile and would complement similar studies in the
Magellanic Clouds. All these areas of the sky have the advantage over the
solar neighbourhood that the objects they contain are at more-or-less known
distances. In this paper, we take the distance modulus of the NGC6522 field
to be 14.7 (Glass et al, 1995). This figure is based on Mira variables in
the nearby Baade's Window field Sgr\,I and takes the d.m. of the LMC
to be 18.55.

%fig 1
\begin{figure}
\epsfxsize=8cm
\epsffile[39 76 573 716]{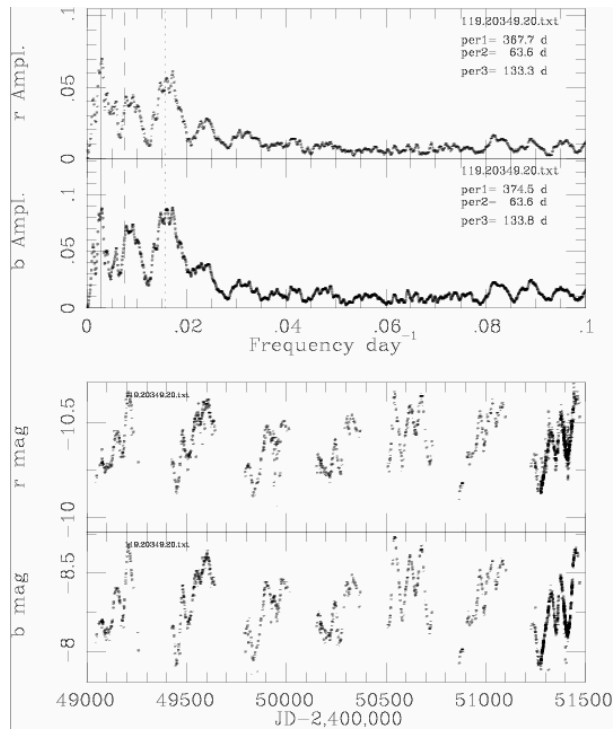}

\caption{Example of MACHO data for the doubly periodic variable
119.20349.20. The ordinates are the semi-amplitudes of the best fitting sine
curves at the frequencies on the abscissas. The lower panels show the light
curves and the upper panels the Fourier transforms for each of the two
colours. This star was classified as having short periods of 63.6d and 133d,
long period of 368d, average (full) amplitudes (from the light curve) of
short period(s) 0.2 mag and of long period 0.3 mag. Its secondary short
period, at just over twice the primary short period, is relatively long and
could be the first harmonic. Diagrams of this type were prepared for every
star. The period with the highest amplitude is shown as a solid line, the
next as a dashed line and the third as a dotted line. Note that the Fourier
spectra have been blocked (i.e. running averaged) by 3 channels, which
accounts for the fact that the period lines shown do not always correspond
to the highest peaks.}

\end{figure}

\section{The data}

The MACHO catalogue for stars lying between RA = 18h 02m 09s and 18h 05m 02s
and Declination -30$^{\circ}$ 16$''$ 39$'$ and -29$^{\circ}$ 46$''$ 43$'$,
(2000) in MACHO field 119, corresponding roughly to the area searched
photographically for large-amplitude variables by Lloyd Evans (1976), was
downloaded and cross-correlated with the DENIS catalogue, limiting the
choice to stars brighter than $K$ = 9.75, fainter than which the work of GS
suggests that relatively few variables will be found. A total of 1661 stars
were selected. The search radius was 3$''$ and the mean distance between
objects was 0.8$''$, with rms scatter 0.4$''$

There were 153 stars with $K_S$ $<$ 9.75 that seemed to have no MACHO
counterparts. Some of these fell on the gaps between CCD chips in MACHO or
off one edge of the field (due to a slight mis-match of boundaries). Of the
remaining 44, seven stars fell in the region of the clusters NGC\,6522 and
NGC\,6528 that were probably too crowded for the MACHO reduction process.
The other 37 appear to have saturated the MACHO CCDs; their positions are
surrounded by areas free of stars for several arcsec radius. For example,
one is coincident with the Lloyd Evans (1976) Mira D9. Stars of earlier
spectral type than M have smaller $R-K$ and are therefore more likely to
saturate the MACHO detectors - see also figs 2 and 3).  

The detailed data for each of the 1661 stars were downloaded from the MACHO
website. The red and blue light curves as well as the frequency power
spectrum of each star in each colour were plotted, in batches of 20. The
range of frequencies searched was from 0 to 0.1 cycles per day (10 days to
infinity in period). The three most conspicuous maxima in each spectrum were
extracted automatically, but each light curve and frequency spectrum was
also examined by eye. In most cases, the automatically extracted periods
agreed between the two colours. There were a significant number of cases
where there was no MACHO $r$ coverage due to a failed CCD. Occasionally, the
data were so sparse as to be useless in one or both colours. The MACHO
quality flags (Alcock et al, 1999) were usually disregarded as they were
often found to exclude useful data. Instead, the Fourier transform programme
was relied upon (see below) to exclude discrepant points. The limit of
detectability is estimated to be two or three hundredths of a mag full
amplitude.

Multiple periodicity is quite prevalent and a distinction is made between
{\bf doubly-periodic} variables, which show a long period (usually a few
hundred days) superimposed on a short period, usually some tens of days, and
{\bf stars with secondary short periods}, where both periods are usually
tens of days (see also section 7.1). Fig 1 shows an example of a
typical classification, in this case a doubly-periodic variable which also
has a secondary short period, though one that is anomalously long.

Of the 1661 selected stars, 1085 were found to be variables and 385 were not
variable. The remainder consisted of 58 that were clearly saturated at both
MACHO $b$ and $r$ and 133 whose data were sparse or negligible in both
wavelengths. The last category were usually, in fact, saturated but in a
minority of cases were probably crowded.

We did not find any variables outside the categories of semi-regulars and
Miras.

\begin{table*}
\begin{minipage}{16.5cm}
\caption{Lloyd Evans (1976) stars and other large-amplitude variables}
\begin{tabular}{llllllllllll}
TLE &  $P_{LE}^1$ & $P_{MA}^2$ & Amp$^3$ &  MACHO & comments$^4$ & TLE &  $P_{LE}$ & $P_{MA}$ & Amp$^1$ &  MACHO & comments$^2$\\
name & day & day & mag & 119. &                        &  name & day & day & mag & 119. & \\   
 395 & 115 &  116 &  2.5 & 20091.28  &                  &  238 & 290:&      &      &           &  5$''$ dead zone \\
 440 & 117 &  116 &  3.0 & 19700.529 &                  &  D3  & 305 &      &      & 20351.6844&  sat, $b$ only   \\
 792 & 126 &  124 &  1.0 & 20349.17  &  SRV             &  786 & 305 &  320 &  3.5 & 20610.46  &            \\
 181 & 180 &  185 &  2.6 & 19833.209 &                  &  207 & 330 &  317 &      & 19701.6529&  sat, no $r$ * \\
 574 & 180 &  185 &  3.3 & 20738.63  &                  &  A29 & 335 &  339 &      & 20743.4738&  sparse, no $r$ \\
 51  & 190 &  198 &  5.0 & 20224.22  &  $b$ only        &  403 & 335 &  340 &  3.5 & 20090.355 &             \\               
 721 & 190 &  195 &  4.0 & 20353.109 &  see note 5      &  A3  & 340 &  345 &  4.0 & 20484.176 &             \\
 313 & 210 &  218 &  3.1 & 20742.1740&                  &  120 & 345 &      &  4.0 & 20613.5347&  sat, no $r$  \\
 652 & 230 &  237 &  3.0 & 19958.30  &                  &  830 & 360 &      &      &           &  6$''$ dead zone \\
 320 & 240 &  240 &  4.0 & 20611.55  &                  &  D6  & 360 &  376 &  3.0 & 20611.462 &             \\
 791 & 240 &  240 &  2.5 & 20219.212 &                  &  590 & 375 &  186 &  3.5 & 20478.54  &             \\
 200 & 250 &      &      &           &  5$''$ dead zone &  D5  & 400:&  446 &  3.5 & 20483.26  &  *          \\
 826 & 250 &  265 &  4.0 & 20743.538 &  $b$ only        &  D9  & 400:&      &      &           &  see note 6 \\
 A28 & 260 &      &      & 19698.251 &  SRV 2 pers, *   &  D11 & 420:& 525: &      & 20484.5561&  see note 7 \\
 745 & 260 &  246 &  3.4 & 20223.6720&  $b$ only        &  434 & 420:&  411 &  2.0 & 19700.93  &             \\
 796 & 260 &  285 &  4.0 & 19829.124 &                &  644 & 430 &  413 &  3.0 & 20088.3574&             \\
 D1  & 265 &  257 &  2.0 & 20094.5849&                  &  103 & 465 &  407 &  4.0 & 20744.1387&  blue only; sat \\
 136 & 270 &  317 &  3.2 & 20352.38  &                  &  426 & 465 &  453 &  2.5 & 19830.33  &             \\
 228 & 270 &      &      &           &  15$''$ dead zone&  435 & 465 &  525 &  3.5 & 19830.34  &             \\        
 332 & 270 &  294 &  3.0 & 20741.325 &                  &  205 & 470:&  531 &  3.0 & 19702.2787&             \\
 340 & 270 &  282 &  3.5 & 20610.174 &\\
 336 & 270 &  266 &  3.0 & 20610.39  &                  &      &     &  144 & 1.7  & 20481.28  &  \\
 A5  & 275 &  298 &      & 20742.821 &  $b$ only *      &      &     &  349 & 2.0  & 20741.923 &  \\
 575 & 280 &  279 &  4.0 & 20608.43  &\\                  

\end{tabular}

\vspace{1mm}

Notes: `sat' denotes saturated. $^1LE$ = Lloyd Evans. $^2MA$ = MACHO.
$^3$Average amplitudes are given. $^4$The radii of the dead zones are given
in arcsec. $^5$TLE721 is also listed by MACHO as 119.20353.111 which has
only got a red curve. The positions are almost identical. $^6$The image of
D9 is overwhelmed by a nearby bright star. $^7$This object is sparsely
covered, very faint and falls in an area with only blue MACHO coverage. It
shows a period of about 525d but an amplitude of $\leq$ 1 mag, which may be
untrustworthy. *See text for explanation.

\end{minipage}
\end{table*}

\subsection{Interpretation of the Fourier power spectra}

The Fourier transform programme removes iteratively all data more than 2.58
standard deviations from the mean and searches for the three most prominent
amplitudes using prewhitening between each step.

Some spectra, later disregarded, were apparently distorted by saturation and
gave false frequency maxima. The light `curves' in question are usually fairly
flat, but with erratic excursions up or down, and the data points were
usually near the top of the range in $r$ and $b$, but not uniformly so. In
some cases, only one of $r$, $b$ are saturated and the variations can be
seen in the other.

Very often the most prominent feature of the Fourier spectrum is a
moderately broad peak with a spike on top. The programme picks out the
highest peak - the spike. This can be due to noise, but is more likely a
manifestation of a period of highly regular behaviour. The position of the
spike was usually close to the maximum of the broad peak. Sometimes the blue
and red spikes did not coincide, but the maximum of the broad peaks agreed
within a few hundredths of a cycle per day. The position of the red spike
was usually the value noted. The periods were normally recorded to three
significant figures.

The breadth of the peak is largely determined by the spacing of the
observations (the `window function') but can also be affected by
irregularity of frequency and amplitude.

Noted were the most prominent short period, the second most prominent short
period and the most prominent long period (see fig 1). The long period was
sometimes not visually apparent from the light curves but appeared in both
the red and blue Fourier spectra. In such cases, the amplitude was very
small and noted as 0.05 mag.

The full short-period amplitude was estimated by eye from the $r$ light
curve whenever possible, and represents the eye-averaged amplitude at the
most prominent short period or periods. The values listed were 0.05, 0.1,
0.15, 0.2, 0.3, .... mag. Very small but apparently real amplitudes were
classed as 0.05 mag.  In the case of doubly periodic stars, the amplitude at
the longer period was estimated in the same way. The overall or long-term
amplitude was often larger than the quantity just described, whether due to
genuine long periods, apparently random variations in amplitude or long-term
drifts. As a test of the accuracy and reliability of the amplitude
classification, some were repeated independently after a long time. They
were sometimes found to differ by one place in the list from the earlier
estimates.

We have not attempted to sub-divide the variables by regularity,
such as into the traditional regular, SRa, SRb and irregular classes. There
are undoubted trends from irregularity to regularity with amplitude
and period (Lebzelter, Schultheis \& Melchior, 2002).

\section{Particular types of variables}

\subsection{Mira-like variables}

Amongst the 1085 variables, 34 had $r$ amplitudes above 1.5 mag. All but two
of these were Miras or SRVs that had been found previously by Lloyd Evans (1976).
The other two were 119.20741.923, which resembles a Mira with 349d period
but has an $r$ amplitude of only 2 mag, and 119.20481.28, which has a period of
144 days and a $b$ amplitude of 1.7 mag.

There are 44 stars in Lloyd Evans' list (see table 1), of which 10 did not
appear in the DENIS-MACHO cross list of 1661. Of these, one was a non-Mira,
making 9 that should have been there but were not. Six of the 10 were burnt
out in the MACHO images and the remaining four, denoted by asterisks, were
omitted from the cross-correlation because they appear to lack DENIS
entries (two because of saturation; the other two, TLEA28, a semi-regular,
and TLE207, a star with a thick dust shell, were seen in 2MASS). These four
had previously been found in the MACHO data by entering positions derived
from the Lloyd Evans charts and the Digitized Sky Survey.  No additional
Miras were discovered.  It is possible that some of the Mira light curves
were saturated at times in MACHO, so that some amplitudes may be truncated.
Table 1 includes two stars (at the end) with $r$ amplitudes of 1.7 and 2.0
mag, which were the only other MACHO objects with amplitudes greater than
1.5 mag. 

The agreement with Lloyd Evans's periods is usually good. However, the
period of his star 590 was found to be 186d and not 375d as previously
determined. This was probably a result of the closeness of the period to
half a year, the seasonality of the observations and the use of photographic
techniques, which are inherently less photometrically precise.

\subsection{Doubly-periodic variables}

Wood et al (1999) found large numbers of double-period stars in the LMC and
similar stars were found in the Baade's Window NGC\,6522 field by Alard et
al (2001). The easiest to find of these stars have short periods of tens of
days, of small amplitude, and a long period of hundreds of days, often with
somewhat higher amplitude.

The proportion of these double-period stars is harder to evaluate in the
NGC\,6522 field than in the Magellanic Clouds because the seasonal nature of
the observations makes the determination of the long periods less reliable,
especially when their amplitudes are small. When all traces of periodic
variation, apparent in both colours, are accepted the number of such stars
is 189. If only those with amplitude $>$ 0.1 mag are counted, it becomes
113. Sources with amplitude $>$ 0.2 mag are 55 in number. The maximum
secondary amplitude observed is 1.1 and usually it is less than 0.5 mag.

The $K$ mags of these variables extend to well below $K$ = 8.2, generally
taken as the upper limit of the RGB (e.g., Tiede, Frogel \& Terndrup, 1995;
Omont et al, 1999)

%fig2
\begin{figure*}
\begin{minipage}{17.5cm}
\epsfxsize=17.5cm
\epsffile[40 133 573 659]{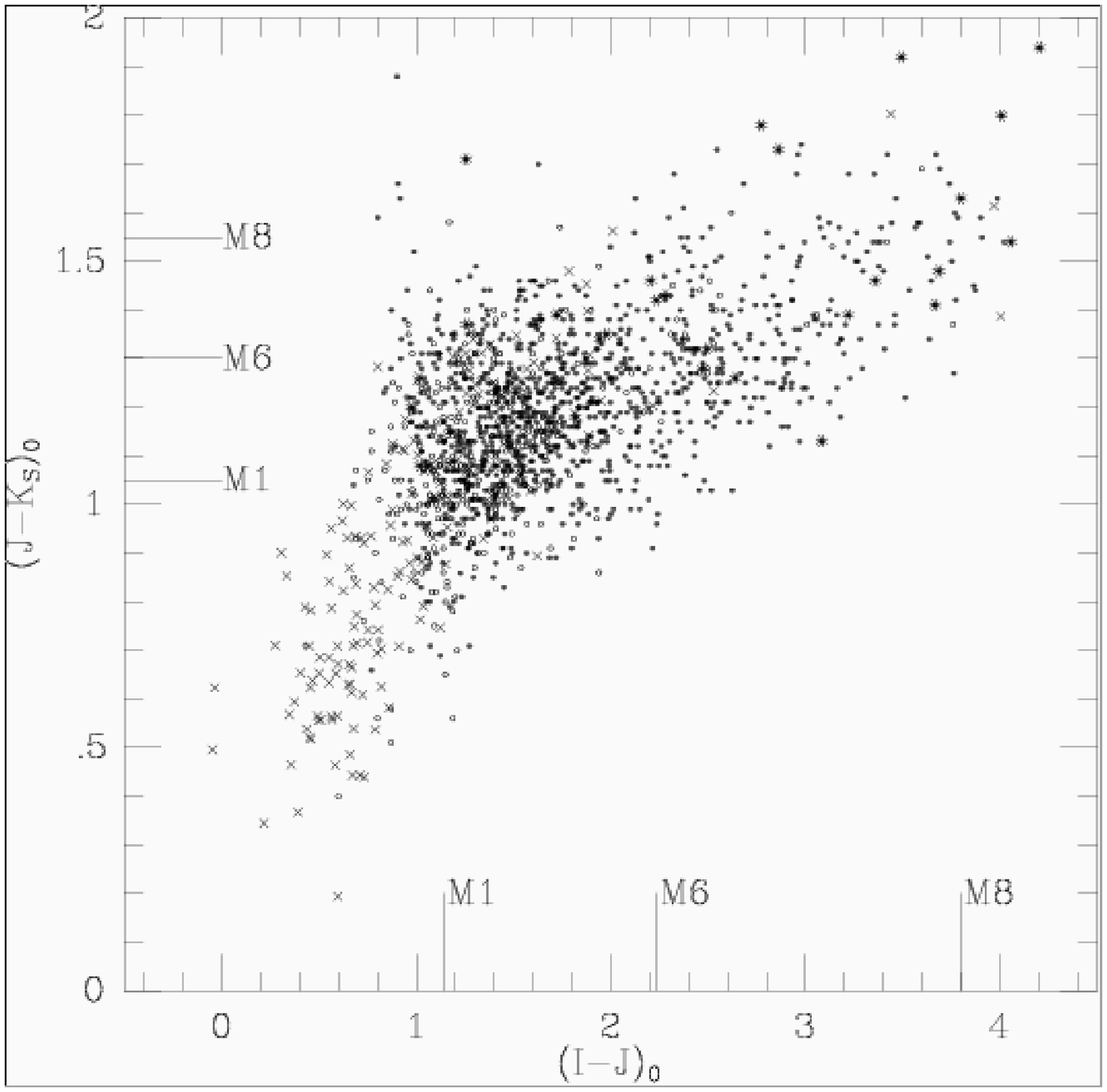}
\caption{$(I-J)$ vs $(J-K_S)$ for the objects in the survey. Crosses are
saturated or have sparse data in MACHO. Asterisks are the large-amplitude
variables of Table 1 (mainly Miras). Solid points are all other variables.
Hollow points are non-variables. The colours of some M-K types are indicated,
based on fig 5 of GS. Most of the objects are of M type, with the earlier
types more frequent than the later. The saturated stars are concentrated
towards significantly earlier types.}
\end{minipage}
\end{figure*}

%fig3
\begin{figure*}
\begin{minipage}{17.5cm}
\epsfxsize=17.5cm
\epsffile[10 132 572 661]{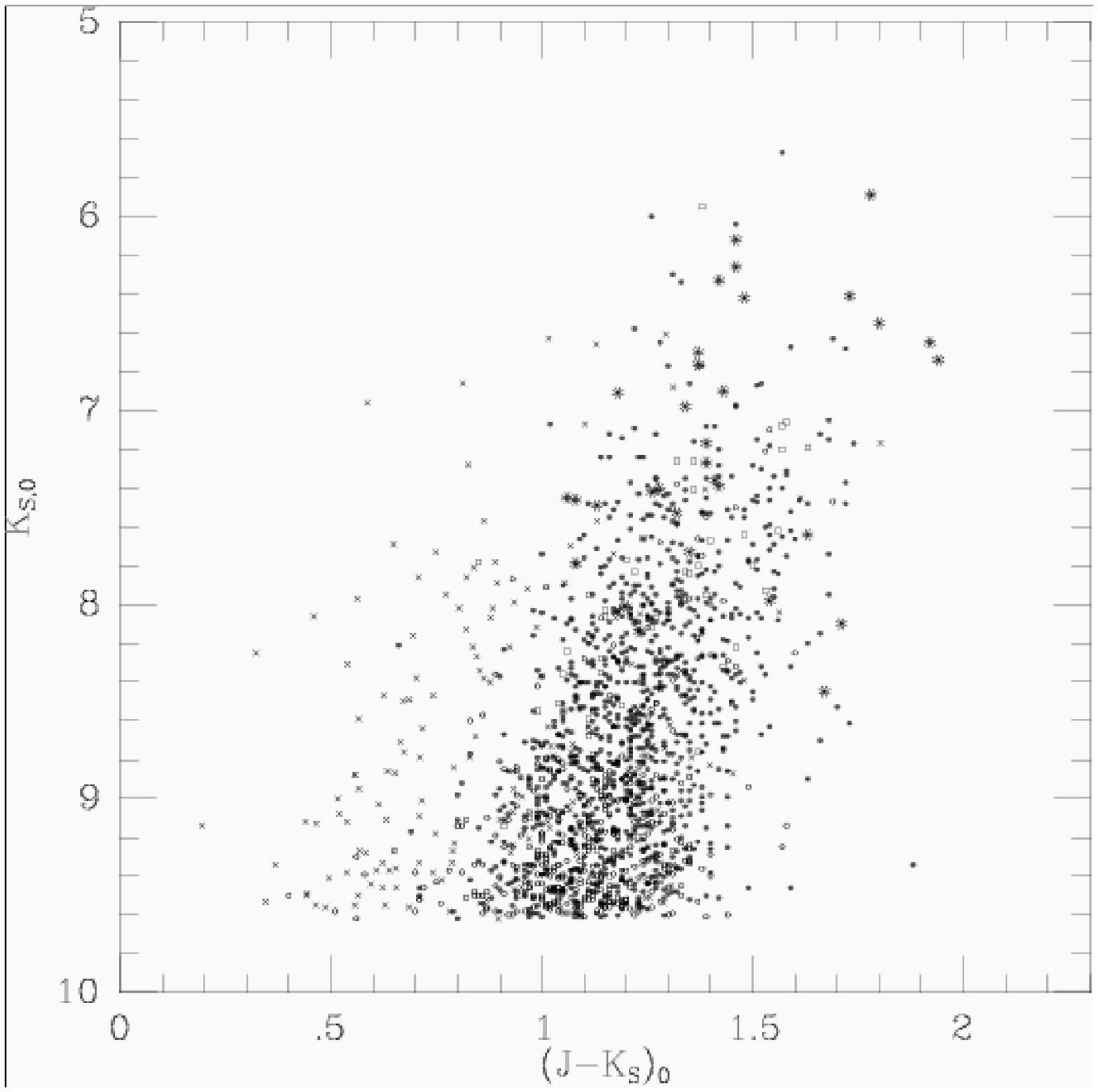} 
\caption{The $K_{S,0}$ vs $(J-K_S)_0$ diagram for all of the sample that
have $J$ and $K_S$ magnitudes. The stars either saturated or having minimal
data in $r$ or $b$ are denoted by crosses, and are clearly for the most part
of earlier spectral type (see also fig 1) than M. The non-variables are
shown as open circles and are concentrated towards fainter $K_0$ as expected
(see GS). The doubly periodic variables are shown as open squares, and are
mixed with the normal SRVs, shown as solid dots, though occurring
more frequently at brighter $K_0$. The large-amplitude variables are shown
as asterisks, and are mostly Miras (but see text).}
\end{minipage}
\end{figure*}

\section{Infrared photometry}

The DENIS photometry shows good agreement at $J$ and $K$ with 2MASS and its
standard errors should be less than 0.10 to 0.12 mag at each wavelength
(Schultheis \& Glass, 2001).

The interstellar absorption $A_V$ within the NGC\,6522 field has been
determined on a star-by-star basis using the programme of Stanek (1996),
corrected as prescribed on the web site that he cites. The reddening law was
taken to be
\[
A_V:A_I:A_J:A_{K,S}= 1:0.59:0.245:0.093
\]
In a few cases, such as near the edges of the field or near the clusters
NGC\,6522 and NGC\,6528, the programme (map.f) could not function and $A_V$
was simply taken to be 1.36 (Alcock et al, 1997).

\subsection{Colour-colour diagram}

The DENIS colour-colour diagram $(I-J)$ vs $(J-K_S)$ (fig 2) of the survey
objects confirms that they are mainly of M spectral type, similar to the
stars examined by GS, to which these are a super-set. The saturated stars
lie blue-ward of the others in both colours and are clearly of hotter
spectral types in general. The figure indicates the colours of some
Morgan-Keenan spectral types, taken from GS.

\subsection{Colour - magnitude diagram}

The $K_{S,0}$, $(J-K_S)_0$ diagram is shown in fig 3. The general features
of this diagram are similar to that presented by Schultheis \& Glass (2001).

Some of the apparent scatter is caused by foreground objects along the line
of sight. In addition, there are many objects whose MACHO observations,
though not their infrared photometry, are saturated. These are denoted by
crosses. These are mostly stars of earlier type than M and are predominantly
located blueward of the variables in $(J-K_S)_0$.

Several types of objects have been distinguished in the figure. As expected,
the Miras (asterisks) are usually towards the bright end of the
$K$-distribution, although some may have dust shells which place them at
fainter $K$ and redward of the main distribution. One variable of fairly
large amplitude, TLE\,395, having a period of only 116 days, is quite faint,
at ($K_{S,0}$, $(J-K_S)_0$) $\sim$ (9.6, 1.1).

The majority of `ordinary' variables are indicated by small solid symbols.
The non-variables are shown as small open circles. The proportion of
variables can be seen to be low at faint $K$ and to increase strongly with
luminosity. This is in accord with the results of GS.

The double-period stars are distinguished by hollow squares. They are evenly
spread amongst the ordinary variables in colour, but are relatively more
numerous at higher luminosities. However, it should be noted that we have
confined the sample to the 55 doubles with long-period amplitudes in excess
of 0.2 mag, in the interest of reliability, and this may have a systematic
effect on their distribution in the figure.

\section{Amplitude distribution}

%fig4
\begin{figure}
\epsfxsize=8cm
\epsffile[28 497 539 786]{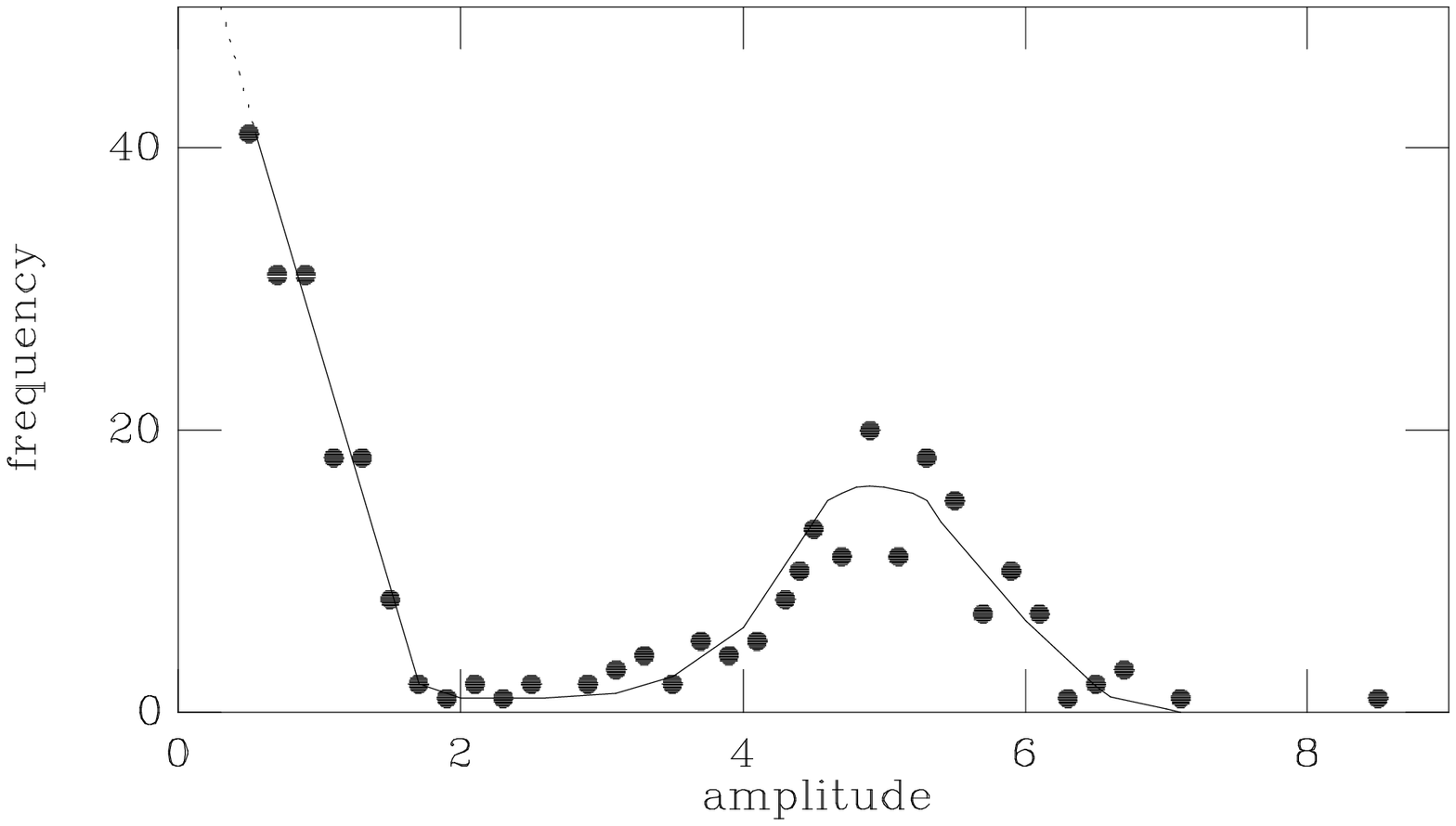}
\caption{Frequency of mean amplitudes of red variables, from
Payne-Gaposchkin, (1951). The amplitudes are presumed to be in the blue
photographic band.}
\end{figure}

%fig5
\begin{figure}
\epsfxsize=8cm
\epsffile[39 256 574 536]{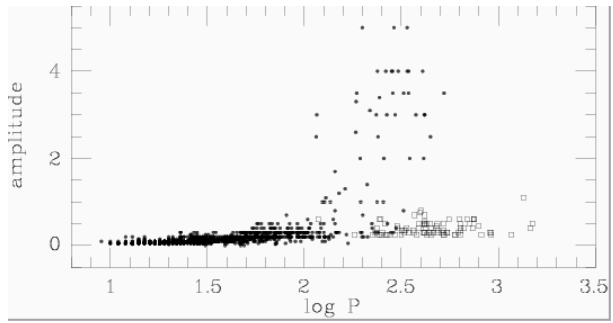}
\caption{ MACHO $r$ amplitudes vs period for the variables. The open points
represent the amplitudes of the longer periods of the doubly-periodic
variables. }
\end{figure}

%fig6
\begin{figure}
\epsfxsize=8cm
\epsffile[28 515 539 786]{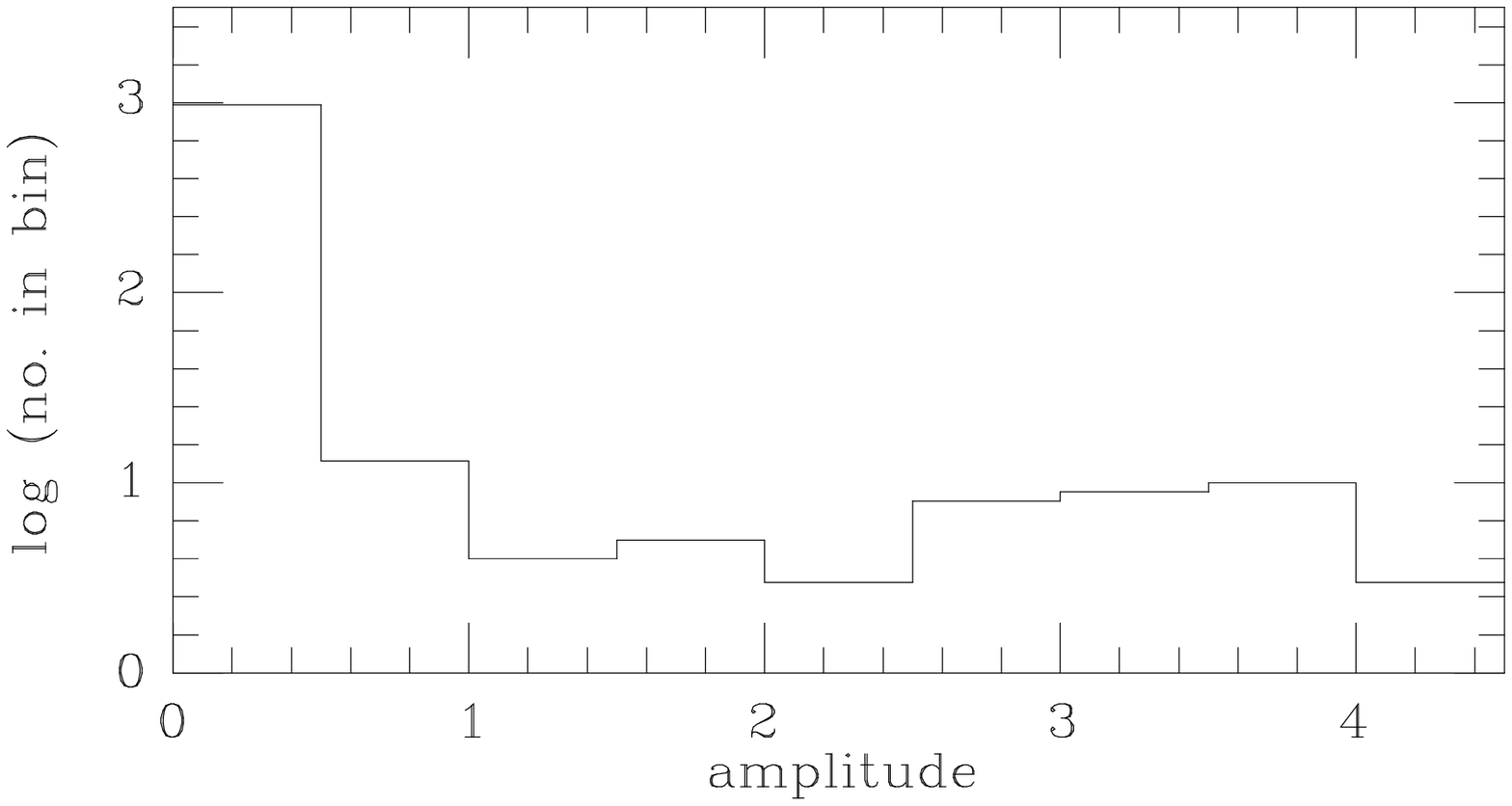}
\caption{Histogram of $r$ amplitudes found for stars in the NGC6522 field.
Note that the primary short-period amplitudes were used.}
\end{figure}

The canonical view of the frequency distribution of the mean amplitude of
red variables was given by Payne-Gaposchkin (1951; see fig.\ 4). The
dividing line between small-amplitude variables and Miras is traditionally
taken at an amplitude of 2.5 mag in $B$ or $V$. At the time, most of the
information would have come from photographic photometry, which is not
precise enough to identify variables with amplitudes of a few tenths of a
mag or less. This is reflected in the dotted line at low amplitudes in her
diagram. There appears to be a marked break in the distribution between 2
and 3 mag in amplitude.

The $r$ amplitude distribution vs period is shown in fig 5.  The most
prominent short period is used in the abscissa. In the case of
large-amplitude variables, such as Miras, the shortest period can be several
hundred days. The longer periods of the double-period stars are denoted by
open symbols. Above a MACHO $r$ amplitude of about 0.6 the number of
variables drops very sharply. Variables with larger amplitudes than this
have periods longer than 200 days. The numbers of small-amplitude variables
drop off at a period of around 150 days but the long periods of
double-period stars continues the small-amplitude sequence to longer
periods.

In figure 6 we show the $r$ amplitude distribution of variables in the
NGC\,6522 field as a histogram with 0.5 mag boxes. In comparing figs 4 and
6, it should be borne in mind that the data in the former are probably of
heterogeneous origin and that $B$ and $V$ amplitudes of late-type variables
are somewhat greater than those at MACHO $r$ because they represent the
short-wavelength tails of the stellar energy distributions and are thus
likely to be more temperature-sensitive. The $r$ amplitudes of the stars in
Table 1 with asterisks have been estimated from the MACHO light curves and
are included in the histogram. The figure is slightly distorted by the
absence of the other 5 Miras, which would probably fall into the bins around
3 mag. Nevertheless, it is clear that the short period stars outnumber the
Miras in the field by about two orders of magnitude (or more if there are
many variables below the amplitude limit of the present survey).

It is also clear that the frequency of moderately large-amplitude variables
in the region between the Miras and the small-amplitude variables is not as
low as in the Payne-Gaposchkin diagram, even though the numbers in each box
are small. This reflects the ease of finding Miras in spite of their rarity,
while the less-conspicuous intermediate types, although only slightly rarer,
were not readily discoverable. The traditional amplitude discriminant for
Miras is thus a somewhat weak one.

\section{$K$ vs log$P$}

\subsection{The $K$ vs log$P$ sequences}

%fig7
\begin{figure*}
\begin{minipage}{17.5cm}
\begin{center}
\epsfxsize=15cm
\epsffile[39 206 573 586]{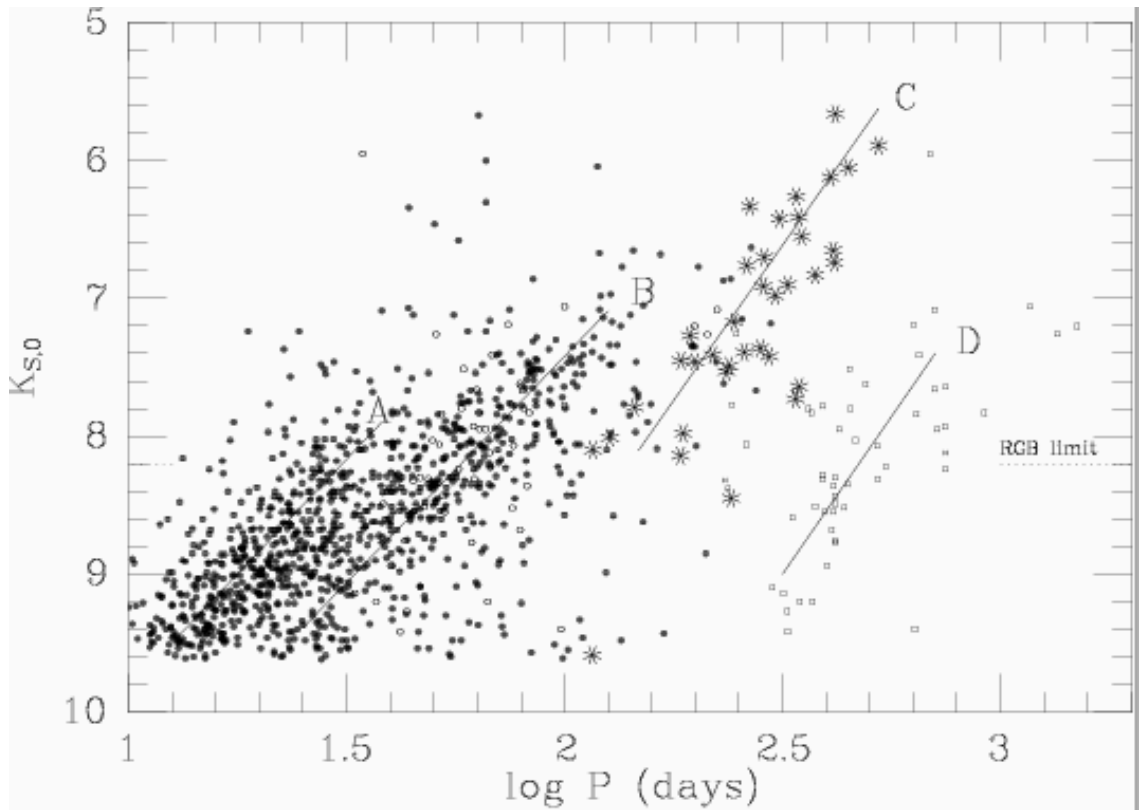} 
\end{center}
\caption{$K$ vs log\,$P$ diagram for the variables found in the Baade's
Window NGC\,6522 field. The general distribution of stars in this diagram
resembles those in the LMC (Wood, 2000; Cioni, 2001) and the SMC (Ita et al,
in preparation). Variables not otherwise distinguished are given as solid
dots. Doubly-periodic variables (as defined in section 2 and having
secondary $r$ amplitudes greater than 0.2 mag), are denoted by open symbols.
Large-amplitude stars (with primary $r$ amplitude $\ge$ 1.5 mag) are shown
as asterisks. Because of the complicated overlaps of the different
sequences, the lines that are shown were fitted by eye to the loci of each
sequence. Amongst the large-amplitude stars, the lowest is TLE\,395, with
semi-regular behaviour (Lloyd Evans, 1976). A small number of objects are
brighter than typical; these may be in the foreground.}
\end{minipage}
\end{figure*}

The $K$, log$P$ diagram of the NGC6522 field, fig 7, bears a considerable
resemblance to that in the LMC (Wood, 2000; Cioni et al, 2001). There are
four distinct sequences visible, denoted A, B, C and D. The freehand lines
on the figure satisfy the following relations, which are given for
convenience only:

\[
K_{S,0}=-3.3\;{\rm log}P +13.1,\; 1.1 < {\rm log}P < 1.6 \;\; \rm{(A)} 
\]
\[
K_{S,0}=-3.3\;{\rm log}P +14.0,\; 1.4 < {\rm log}P < 2.1  \;\;\; \rm{(B)}
\]
\[
K_{S,0}=-4.6\;{\rm log}P +18.1,\; 2.2 <{\rm log} P < 2.7  \;\;\; \rm{(C)} 
\]
\[
K_{S,0}=-4.6\;{\rm log}P +20.4,\ 2.5 < {\rm log}P < 2.9   \;\;\; \rm{(D)}
\]

The rather steep slope of relation C (the Miras) compared to previous work
(e.g., Glass et al, 1995) may be due to the fact that these variables have
large amplitudes and that only single observations at $K_S$ are available. A
very few extra observations could change the slope. The results of Glass et
al (1995) indicate that the Miras in the nearby Baade's Window Sgr I have
the same slope as the LMC $K$, log\,$P$ relation within $\pm$ 0.4 mag per
unit of log $P$.

\begin{itemize}

\item Sequence A contains the shortest period stars, usually with small
amplitudes, and has few members brighter than the upper limit of the Red
Giant Branch (RGB) at $K_0$ $\sim$ 8.2.

\item In Sequence B are found, besides many normal SRVs, the shorter
periods of the doubly-periodic stars, denoted by hollow symbols, whose
longer periods occupy sequence D. Sequence B extends to $K_{S,0}$ $\sim$ 7,
the luminosity level of the 200-day Miras. 

\item Sequence C is that of the large-amplitude variables, having $r$
amplitude $>$ 1.5 mag. Nearly all of these are known Miras, discovered by S.
Gaposchkin and others, as listed by Lloyd Evans (1976). Since the points
shown here are single $K_S$ observations of large-amplitude variables, a
close fit to a period-luminosity relation cannot be expected. The two
outlying stars near log$P$ = 2.5, $K$=7.7, viz TLE\,120 and TLE\,A3, have
been observed by Glass \& Feast (1982), with similar $K$ values but TLE\,A3
has been seen at $K$ = 6.95 by Frogel \& Whitford (1987). The star at log$P$
= 2.4, $K$ = 8.4 is TLE\,791, with no previous photometry. The lowest of
these points, at $K$ = 9.6, is TLE\,395, a variable noted by Lloyd Evans to
have semi-regular behaviour and which appears faint in the ISOGAL 7 and 15
$\mu$m measurements (Glass et al, 1999).

\item Each star in sequence D, as mentioned, has a corresponding point in
sequence B. Only those stars with sequence D $r$ amplitudes greater than 0.2
mag are shown, in the expectation that these data will be the most reliable.
(The remainder are included in the ordinary variables shown with solid
symbols). If the stars with lower $r$ mags are included, their secondary
periods cause the space between sequences C and D to be somewhat more
occupied.

\end{itemize}

Some of the points scattered around the diagram will correspond to stars
that lie in front of the Bulge. 

An approximate value for the scatter about the lines A and B in fig 7 was
determined by accepting only those stars in parallelograms about 0.4 mag in
$K$ above and below the freehand lines shown, and limited in $K$ to the
ranges covered by the lines. It was also assumed that the error is entirely
in $K$. The standard deviations of individual stars from the fits were 0.21
and 0.22 mag, which is smaller than that found for the Mira sequence in the
nearby Sgr\,I Baade's Window by Glass et al (1995), viz 0.35. Although Miras
have moderately large amplitudes in the infrared, each of the points in
Glass et al is the mean of several observations. The probably slightly
greater variation of absorption in the Sgr I Baade's Window may, however,
have a modest effect on the dispersion in the Mira $K$ mags. The estimates
made here for the semi-regular variables are bound to be optimistic,
however, because stars deviating more than $\sim$ 2$\sigma$ were rejected as
being outside the parallelograms.

It is clear from visual inspection that the scatter about the sequences is
less pronounced in the LMC (cf fig 1 of Wood, 2000) than in the NGC\,6522
Window. The finite depth of the Bulge is almost certainly the cause of the
higher values observed.

\subsection{Secondary short periods}

Many of the Fourier spectra show more than one short-period peak, as found
earlier in the LMC by Wood et al (1999), and ambiguity may arise when these
peaks are nearly equal in height. To illustrate this, we plot in fig 8
the log of the second most important short period against the log of the
most important short period for all 443 stars classified as having two or
more short periods. Many points fall along (solid) lines that indicate the
difference in log period is $\pm$0.14, or a factor of about 1.4, at least
for periods up to about 50 days, beyond which it seems to increase. The
density of points suggests that the second most important peaks are somewhat
more often at longer periods than shorter. Thus, stars with almost equal
peaks might be misclassified as to period by a factor of $\sim$ 1.4 too high
or too low, and this might explain some of the inter-sequence stars in fig
7.

A smaller number of objects fall along lines having differences in log
period of $\pm$0.27, or a factor of 1.86. These are potential sources of
confusion because this is approximately the separation of sequences A and B
at a given $K_S$. As a test, stars falling within the parallelogram boxes
centered on sequences A and B (described above) were plotted together with
their second short periods on the $K_{S,0}$, log$P$ diagram, distinguishing
the second periods by special symbols. It is thought that because only about
40\% of these stars have secondary short periods and because ambiguity in
deciding between primary and secondary short periods is not very common, the
effect on fig 7 should not be very noticeable.

Wood et al (1999) discuss multi-periodic stars in the LMC similar to those
mentioned here with more than one short-period. Their fig 4 shows that the
factor of $\sim$ 1.4 can be predicted as the ratio of first to second
overtone periods in the radially pulsating models of Wood \& Sebo (1996),
though the other period ratios cannot be fitted quite so satisfactorily. The
fact that distinct A and B loci exist in figure 7 in spite of co-existing
modes must mean that the ratio between periods is fairly stable. In fig 9
we show the short-period ratios plotted in the same manner as in Wood et al.
Although the diagrams agree in that there is a concentration towards a ratio
of $P_{\rm greater}/P_{\rm lesser}$ = 1.4, there are many points towards the
left of our figure with higher ratios that are absent in that of Wood et al.

%fig8
\begin{figure} 
\epsfxsize=8cm 
\epsffile[41 137 571 656]{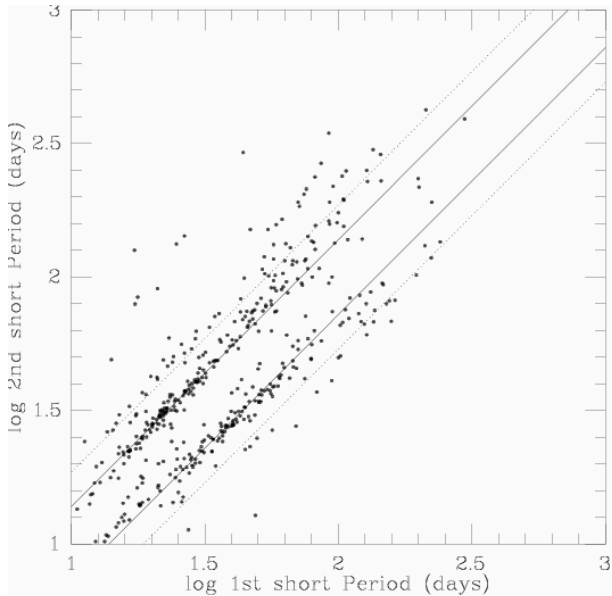} 
\caption{Log of the most conspicuous short period plotted against log of the
second most conspicuous short period. The solid lines represent differences
of $\pm$0.14 between log periods and the dotted lines $\pm$0.27 between log
periods. The first difference seems to be inherent to the variability and
the dotted lines represent the difference between the A and B series.} 
\end{figure}

%fig9
\begin{figure} 
\epsfxsize=8cm 
\epsffile[39 257 573 535]{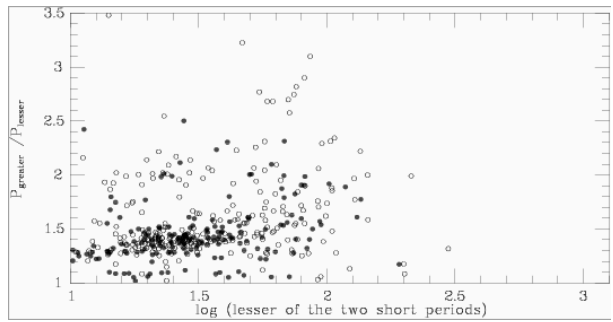} 

\caption{The same data as in fig 8, plotted in the manner of fig 4 of Wood
et al (1999). The open circles are stars in which the shorter of the two
periods has the highest amplitude and the closed circles those for which the
longer of the short periods dominates.}

\end{figure}

\subsection{Solar neighbourhood SRVs}

The named semiregular variables, i.e. those in the {\it Combined General
Catalogue of Variable Stars} (CGCVS) (Kholopov et al, 1998), were mainly
discovered by visual or photographic means and are therefore of moderately
large amplitude, at least when compared to the majority of variables found
here. Knapp et al (2003, in press) show histograms of the period
distribution of late-type variables in the Catalogue (their fig 1). The SRb,
SRa and Mira distributions peak at periods around 100, 150 and 270d
respectively. It is clear that the absence of the small-amplitude variables
distorts the appearances of these distributions.

In the solar neighbourhood, the sample with known parallaxes (and therefore
absolute magnitudes) is small and confined to stars from the CGCVS and
therefore having relatively large amplitude. Since the $K$, log$P$ diagrams
of the LMC and the NGC\,6522 fields show many of the same features, we would
expect a complete $K$, log$P$ diagram of the solar neighbourhood SRVs to be
similar. 

In fact, Bedding \& Zijlstra (1998) have examined a set of
solar neighbourhood SRVs with Hipparcos parallaxes.  The variability
criterion that they used was that the Hipparcos amplitude, $a$, had to
exceed 0.06\,mag.  This amounts to a peak-to-peak variation (the quantity
used in this paper) of 0.12 mag (in fact, most of the Hipparcos light curves
of their stars show considerably higher amplitudes than this). The Hipparcos
photometric band is very broad and has little sensitivity in the red to
near-infrared regions (our variability amplitude was estimated where
possible for the $r$ band of MACHO, and is usually about 75\% of that in the
MACHO $b$ band). If we restrict fig 7 to stars with $r$ amplitude greater
than 0.3 mag,  it is found that most of sequence B below $K$ = 8.5
disappears. This has been done in figure 10.

%fig10
\begin{figure}
\epsfxsize=8.2cm
\epsffile[28 424 539 786]{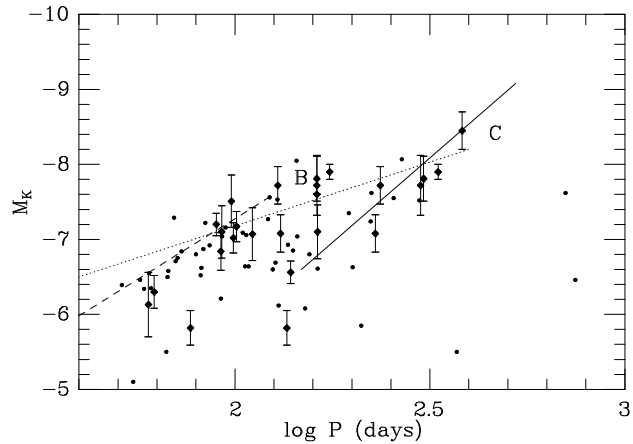}

\caption{The $K$, log$P$ diagram for only those variables from fig 7 with
amplitudes $>$ 0.3 mag and $\leq$ 1.6 mag. Superimposed are the solar
neighbourhood SRVs (diamonds) from Bedding \& Zijlstra (1998). The positions
of the B and C sequences are shown as dashed and solid lines respectively
and the dotted line is these authors' suggested fit to the (upward-adjusted)
evolutionary track that Whitelock (1986), proposed for SRVs in globular
clusters.}

\end{figure}

The Bedding \& Zijlstra stars and the remaining NGC\,6522 large-amplitude SR
variables are seen to occupy rather similar regions of the $K$, log$P$
diagram. This suggests that the dotted line that they draw on their fig 1,
if it has physical significance as an evolutionary track, only applies to
the highest amplitude, most luminous and longest period of the B-sequence SR
variables in the solar neighbourhood, and is not a property of all SR
variables. It remains possible that the whole of the $K_S$, log $P$ diagram
(fig 7, excluding the region of sequence D) represents stars along
evolutionary tracks parallel to those of Bedding \& Zijlstra. Tracks with
the same slope were predicted by Vassiliadis \& Wood (1993).

\subsection{Comparison with LMC}

%fig11
\begin{figure} 
\epsfxsize=8cm 
\epsffile[48 422 539 786]{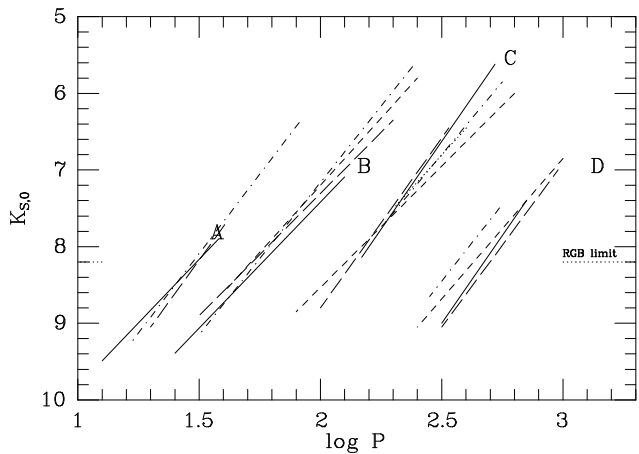} 

\caption{Estimated positions of the sequences in the $K_{S,0}$, log$P$
diagram from this and previous work, reduced to the distance modulus of the
NGC\,6522 field using $\Delta$d.m. = 3.85 for the LMC and 4.35 for the SMC.
The solid lines represent this work and are taken from Fig7.  The Cioni et
al work on the LMC (fig 6 of Cioni et al, 2001) is shown as short dashes.
The Wood (2000) work is shown with long dashes. The Ita et al work (in
preparation) on the SMC is shown as dot-dash lines.}

\end{figure}

Fig 11 compares the estimated positions of the $K_{S,0}$, log$P$ sequences
in the LMC, SMC and NGC6522 fields, after adjusting for distance moduli. 
The data shown have been extracted by different investigators and, because
there is a subjective element in their interpretation, there may be some
systematic errors. The comparison should, therefore, be regarded as
preliminary. Nevertheless, the sequences agree quite well between the
various sets of data, in spite of the known differences in metallicity
between fields. 

The most apparent differences that can be seen relate to the period ranges
that each sequence covers:

\begin{itemize}

\item The A sequence seems to reach to higher luminosities in the SMC and
LMC than in NGC\,6522, though they are few in number above the estimated
tip of the RGB ($K$ $\sim$ 8.2)

\item The B sequence seems to extend to higher luminosities in the
Magellanic Clouds than in the NGC\,6522 field, corresponding to $K_{S,0}$
$\sim$ 6.4, which is comparable to the 400-day Miras. However, the
extension consists mainly of carbon stars, which do not occur in the
NGC\,6522 field.

\item In the LMC data the low-luminosity end of sequence B is populated as
densely as that of sequence A. However, in NGC\,6522 it appears to be much
more sparse.

\item In the LMC, sequence C (the Miras) consists mainly of C stars at
higher luminosities. This range is occupied by M stars in the NGC\,6522
field.

\item The scatter in sequence D is high in the present sample, mainly
because of the seasonal interruptions. The data are probably compatible with
those from the Magellanic Clouds.

\end{itemize}

There are other noticeable ways in which the galactic and Magellanic Cloud M
giants differ. For example, in comparing the infrared $JHK$ colours of Miras
in various period groups, Glass et al (1995) have found that those in the
Sgr I field and solar neighbourhood tend to be redder than similar objects
in the LMC, especially at longer periods.

Lebzelter et al (2002) show that there are also differences in the
distribution of $I-J$ colour with period between the Bulge SRVs and those in
the LMC, in that the Bulge stars show a much larger colour range. This is
almost certainly a result of higher metallicity in the Bulge.

Two other manifestations of the metallicity difference are the high
prevalence of S and C stars in the LMC and the fact that the spectral types
of the Mira variables are much earlier on average than in the Galaxy (see,
e.g., Glass \& Lloyd Evans, 2003, to be published). This is because of their
lower metallicity, consequent lower opacity and hence higher effective
temperature. Nevertheless, the most extreme AGB stars do become as cold as
those in the Bulge, with dust-enshrouded AGB stars in both the LMC (van Loon
et al. 1998) and SMC (Groenewegen \& Blommaert 1998) reaching spectral types
as late as M9 and M10.

\section{Conclusions}

In our survey of the NGC\,6522 clear window of Baade for stars with $K$
brighter than 9.75. Some 1661 objects were found with MACHO counterparts
and, of these, 1085 were found to be variable.

It has been shown that the $K$, log$P$ sequences visible among the late-type
giants in the Magellanic Clouds are also to be seen in galactic stars and
that they are not significantly displaced in position on an absolute
magnitude basis. Thus, in a galaxy where a sufficient number of objects can
be monitored, these objects may have value as distance indicators.

Double period stars that appear in the B and D sequences constitute 10\% to
20\% of the variables. Their long periods rarely have amplitudes above 0.5
mag. They are similar to other variables in the B sequence in colour but are
found predominantly amongst the longer period stars, bright in $K$. They
differ from their Magellanic Cloud counterparts with lower metallicity in
that they have a cut-off at about the limit of the LMC oxygen-rich stars and
thus do not overlap the region of the carbon stars.

The Mira sequence (C) however, extends to $K$ luminosities similar to its
counterpart in the LMC, but the higher-luminosity stars are O-rich rather
than C-rich.  

Because our method for finding variables is less biased than the traditional
ones, we have been able to show that the distribution of $r$ amplitudes of
red variables does not have a conspicuous gap between the large-amplitude
variables (Miras) and the semi-regular variables. 

\section{Acknowledgments}

We thank Dr Chris Koen, South African Astronomical Observatory, for help and
advice concerning the Fourier transformation programs. Dr T. Lloyd Evans
made helpful comments on a draft of the paper. Mr Y. Ita and collaborators
gave permission to quote their work.

ISG thanks the Institut d'Astrophysique de Paris for their hospitality
during part of this work. This visit was supported by the CNRS/NRF
agreement.

MS is supported by the APART programme of the Austrian Academy of Science.

This paper utilizes public domain data originally obtained by the MACHO
Project, whose work was performed under the joint auspices of the U.S.
Department of Energy, National Nuclear Security Administration by the
University of California, Lawrence Livermore National Laboratory under
contract No. W-7405-Eng-48, the National Science Foundation through the
Center for Particle Astrophysics of the University of California under
cooperative agreement AST-8809616, and the Mount Stromlo and Siding Spring
Observatory, part of the Australian National University.

We acknowledge use of the Digitized Sky Survey produced at the Space
Telescope Science Institute under US Government grant NAG W-2166, based on
material taken at the UK Schmidt Telescope, operated by the Royal
Observatory Edinburgh with funding from the UK Science and Engineering
Research Council and later by the Anglo-Australian Observatory.


\begin{thebibliography}{}

\bibitem[]{}Alard C. et al, 2001, ApJ, 552, 289

\bibitem[]{}Alcock C. et al, 1997, ApJ, 494, 396

\bibitem[]{}Alcock C. et al, 1999, PASP, 111, 1538

\bibitem[]{}Bedding T.R, Zijlstra A.A., 1998, ApJ, 506, L47

\bibitem[]{}Cioni M.R., Marquette J.-B., Loup C., Azzopardi M., Habing H.J.,
Laserre T., Lesquoy E., 2001, A\&A, 377, 945

\bibitem[]{}Epchtein N., 1998, in Epchtein N., ed., The Impact of Large Scale
Near-IR Sky Surveys, Kluwer, Dordrecht, p. 3

\bibitem[]{}Frogel J.A., Whitford A.E., 1987, ApJ, 320, 199

\bibitem[]{}Glass I.S., Alves D., the ISOGAL and MACHO teams, 2000, in Lemke
D., Stickel M., Wilke K., eds, Lecture Notes in Physics Ser. 548, ISO
Surveys of a Dusty Universe. Berlin, Springer, p. 363

\bibitem[]{}Glass I.S. et al, 1999, A\&A, 308, 127

\bibitem[]{}Glass I.S., Feast M.W., 1982, MNRAS, 198, 199

\bibitem[]{}Glass I.S., Lloyd Evans, 1981, Nature, 291, 303

\bibitem[]{}Glass I.S., Lloyd Evans, 2003, MNRAS in press

\bibitem[]{}Glass I.S., Schultheis M., 2002, MNRAS, 337, 519 (GS)

\bibitem[]{}Glass I.S., Whitelock, P.A., Catchpole R.M., Feast M.W., 1995,
MNRAS, 273, 383

\bibitem[]{}Groenewegen M.A.T., Blommaert J.A.D.L., 1998, A\&A, 332, 25

\bibitem[]{}Ita Y., Tanab\'{e} T., Matsunaga N., Nakajima Y., Nagashima C.,
Nagayama T., Kato D., Nagata T., Nakada Y., 2003, MNRAS, submitted

\bibitem[]{}Kholopov P.N., Samus N., Frolov M.S. et al (eds), 1998, Combined
General Catalogue of Variable Stars.

\bibitem[]{}Knapp G.R., Pourbaix D., Platais I., Jorissen A., 2003, A\&A, in
press

\bibitem[]{}Lebzelter T., Schultheis M., Melchior A.L., 2002, A\&A, 393,
573, 2002

\bibitem[]{}Lloyd Evans T., 1976, MNRAS, 174, 169

\bibitem[]{}Omont A., et al 1999, A\&A, 348, 755

\bibitem[]{}Omont A., et al 2003, A\&A, in press

\bibitem[]{}Payne-Gaposchkin C., 1951, in Astrophysics, ed Hynek J.A.,
McGraw-Hill, p.522
   
\bibitem[]{}Schultheis M., Glass I.S., 2001, MNRAS, 327, 1193

\bibitem[]{}Skrutskie M.F., 1998, in Epchtein N., ed., The Impact of Large
Scale Near-IR Sky Surveys, Kluwer, Dordrecht, p. 11

\bibitem[]{}Stanek K.Z., 1996, ApJ, 460, L37

\bibitem[]{}Tiede G.P., Frogel J.A., Terndrup D.M., 1995, AJ, 110, 2788

\bibitem[]{}van Loon J. Th. et al, 1998, A\&A, 329, 169

\bibitem[]{}Vassiliadis E., Wood P.R., 1993, ApJ, 413, 641

\bibitem[]{}Whitelock P.A., 1986, MNRAS, 219, 525

\bibitem[]{}Wood P.R., 2000, Publ. Astr. Soc. Australia, 18, 18

\bibitem[]{}Wood P.R. and the MACHO team, 1999, in Asymptotic Giant Branch
Stars, IAU Symp. 191, eds. Le Bertre T., L\`{e}bre A., Waelkens C., p. 151

\bibitem[]{}Wood P.R., Sebo K.M., 1996, MNRAS, 282, 958 

\end{thebibliography}
\end{document}